\documentclass{emulateapj}

\slugcomment{ApJ in press}

\shorttitle{Electron Acceleration at Low-Mach-Number Shocks}
\shortauthors{Umeda, Yamao \& Yamazaki}

\begin{document}

\title{
Electron acceleration at
a low-Mach-number perpendicular collisionless shock
}

\author{Takayuki Umeda and Masahiro Yamao}
\affil{Solar-Terrestrial Environment Laboratory, Nagoya University, 
Nagoya, 464-8601, Japan; umeda@stelab.nagoya-u.ac.jp}
\and
\author{Ryo Yamazaki}
\affil{Department of Physical Science, Hiroshima University,
Higashi-Hiroshima 739-8526, Japan; ryo@theo.phys.sci.hiroshima-u.ac.jp}
%


\begin{abstract}
A full particle simulation study is carried out 
on the electron acceleration at a collisionless, 
relatively low Alfven Mach number ($M_A=5$), perpendicular shock. 
Recent self-consistent hybrid shock simulations have demonstrated 
that the shock front of perpendicular shocks 
has a dynamic rippled character along the shock surface 
of low-Mach-number perpendicular shocks. 
In this paper, 
the effect of the rippling of perpendicular shocks 
on the electron acceleration 
is examined by means of large-scale 
(ion-scale) two-dimensional full particle simulations. 
It has been shown that 
a large-amplitude electric field is excited at the shock front 
in association with the ion-scale rippling, 
and that reflected ions are accelerated upstream 
at a localized region where the shock-normal electric field 
of the rippled structure is polarized upstream. 
The current-driven instability caused by the highly-accelerated 
reflected ions has a high growth rate 
to large-amplitude electrostatic waves. 
Energetic electrons are then generated 
by the large-amplitude electrostatic waves 
via electron surfing acceleration 
at the leading edge of the shock transition region. 
The present result suggests that 
the electron surfing acceleration is also a common feature 
at low-Mach-number perpendicular collisionless shocks. 
\end{abstract}



\section{Introduction}

Collisionless shocks are universal processes in space and
are observed in laboratory, astrophysical, and space plasmas,
including astrophysical jets, an interstellar medium,
the heliosphere, and planetary magnetospheres.
Particle acceleration is 
a common feature of at collisionless shocks, 
but is still a major unresolved issue. 
The most plausible mechanism of the acceleration 
is the diffusive shock acceleration (DSA), 
which explains broadband power-law spectrum with an index around 2 
\citep{drury1983,blandford1987}. 
Before the DSA phase in which electrons cross 
the shock front many times, 
they have to be pre-accelerated by unknown injection mechanism.
Such ``injection problem'' is still unresolved.

One of the possible injection mechanisms for electrons 
is the electron surfing acceleration by electrostatic fields 
\citep{Katsouleas_1983,Shimada_2000,Hoshino_2002}. 
\citet{Shimada_2000} performed 
a one-dimensional (1D) electromagnetic full particle simulation
and found 
the formation of large-amplitude electrostatic solitary structures 
during the cyclic reformation of 
a high-Mach-number perpendicular shock in a low-beta and 
weakly magnetized plasma. 
The coherent solitary structures 
trap electrons in their electrostatic potential well, 
resulting in significant surfing acceleration of electrons.

So far, the electron surfing acceleration has been argued
for high-Mach-number shocks, and studied by many authors 
\citep[e.g.,][]{McClements_2001,Hoshino_2002,
Schmitz_2002a,Schmitz_2002b,Amano_2007}.
A typical astrophysical example of such high-Mach-number shocks 
is the young supernova remnant (SNR), 
which is thought to be an accelerator
of high-energy electrons, 
emitting bright synchrotron radiation in radio and X-ray bands 
at shocks with Mach number of ${\cal M}_A\gtrsim10^2$
\citep[e.g.,][]{koyama1995,bamba2003,bamba2005}.
However, in many astrophysical environments, 
the electron acceleration at lower-Mach-number 
(${\cal M}_A\lesssim10^2$) shocks is inferred from observations and/or
expected theoretically. 
Possible examples are old SNRs or old SNRs interacting 
with giant molecular clouds 
\citep{claussen1997,chevalier1999,bykov2000,yamazaki2006,brogan2006},
merging galaxy clusters \citep{markevitch2002}, 
AGN outbursts at the galaxy clusters \citep{fujita2007},
and so on. 
The electron acceleration is also observed at Earth's 
bow shock even with a low Mach number of 
${\cal M}_A\sim6.4$ \citep{oka2006}. 
Hence it is interesting to ask 
whether the electron surfing acceleration 
can occur at lower-Mach-number shocks or not.

It has been demonstrated by one-dimensional full particle simulations 
that the electron surfing acceleration takes place 
at a relatively-high-Mach-number shock of 
$M_A\sim10$
but not at a very-low-Mach-number shock of $M_A\sim3$ 
\citep[e.g.,][]{Shimada_2000,Schmitz_2002a}. 
\footnote{
Note that in the full particle simulations, 
a reduced ion-to-electron mass ratio of $m_i/m_e=20$--25, 
is usually adopted for computational efficiency. 
In general, different mass ratio sometimes changes 
the simulation results qualitatively, 
and the simulation results with a reduced mass ratio 
cannot be directly compared with 
astrophysical phenomena or results of 
hybrid simulations. 
Therefore, to avoid a confusion, 
we use a notation $M_A$ 
for the Mach number of shocks with 
a reduced mass ratio, 
while we use another notation ${\cal M}_A$ 
for the real mass ratio. 
}
However, these previous works on the electron surfing acceleration are 
mainly based on 1D simulations, 
in which configuration of shock magnetic fields 
cannot be modified.

It is well known from resuts of 
two-dimensional hybrid-code simulations 
that there exist large-amplitude fluctuations
in the magnetic field and density of the shock transition region 
of lower-Mach-number shocks (${\cal M}_A=5\sim10$) 
\citep{Winske_1988}. 
These fluctuations at the shock surface 
have been analyzed in terms of 
turbulent ``ripples'' \citep{Lowe_2003,Burgess_2006a}. 
\citet{Burgess_2006b} has demonstrated, using a combination of 
self-consistent hybrid simulation and test particle calculation, 
that energetic electrons are produced by both magnetic mirroring 
by the rippled structure and by stochastic acceleration 
by magnetic fluctuations keeps particles within the
shock transition region. 
However, electron-scale microscopic instabilities, i.e., 
current driven instabilities are neglected 
in the combination of hybrid simulation and test-particle calculation. 
Hence multi-dimensional full particle simulations are necessary 
for studying the effect of ion-scale rippling of a perpendicular shock 
on the electron surfing acceleration.

The purpose of this paper is to examine 
the effect of the dynamic rippled structuring of the shock front 
on electron acceleration processes 
by the first-principle full particle simulation. 
In order to take into account the rippling of a
perpendicular shock,
simulation domain is taken to be larger than 
the ion inertial length. 

The paper is organized as follows.
Section 2 describes the model and the parameters 
of the full particle simulation.
Section 3 demonstrates the ion-scale structure of a perpendicular shock
found in the full particle simulations and 
the resulting electron acceleration.
Finally, section 4 gives summary of this paper.

\section{Full Particle Simulation}

In this paper, 
a collisionless shock is excited with 
the ``relaxation method'' 
\citep[e.g.,][]{Leroy_1981,Leroy_1982,Umeda_2006,Umeda_2008} 
in which the simulation domain is taken 
in the rest frame of the excited shock. 
Generally speaking, it is not easy 
to perform a large-scale (ion-scale) multi-dimensional 
full particle simulations of collisionless shocks 
even with present-day supercomputers. 
This is because a shock wave excited by conventional method 
becomes unsteady relative to the simulation domain. 
Very recently, however, 
a two-dimensional shock-rest-frame model has been successfully 
developed by \citet{Umeda_2008}, 
which is important 
to be able to follow the shock for a long time 
with a limited computer resource.

We use a 2D electromagnetic particle code 
\citep{Umeda_PhD}, in which 
the full set of Maxwell's equations and 
the relativistic equation of motion for individual electrons and ions 
are solved in a self-consistent mannar. 
The continuity equation for charge is also 
solved to compute the exact current density 
given by the motion of charged particles \citep{Umeda_2003}.

The initial state consists of two uniform regions 
separated by a discontinuity. 
In the upstream region that is taken in the left hand side 
of the simulation domain, 
electrons and ions are distributed uniformly in space and 
are given random velocities $(v_x,v_y,v_z)$ to approximate 
shifted Maxwellian momentum distributions 
with the drift velocity $u_{x1}$, 
number density $n_{1} \equiv \epsilon_0 m_e \omega_{pe1}^2 / e^2$, 
isotropic temperatures $T_{e1} \equiv m_e v_{te1}^2$ and 
$T_{i1} \equiv m_i v_{ti1}^2$, 
where $m$, $e$, $\omega_{p}$ and $v_{t}$ are 
the mass, charge, plasma frequency and 
thermal velocity, respectively. 
Subscripts ``1'' and ``2'' denote 
``upstream'' and ``downstream'', respectively.
The upstream magnetic field $B_{y01} \equiv -m_e \omega_{ce1}/e$ 
is also assumed to be uniform, where $\omega_{c}$ 
is the cyclotron frequency (with sign included). 
The downstream region taken in the right-hand side 
of the simulation domain is prepared similarly with 
the drift velocity $u_{x2}$, density $n_{2}$, 
isotropic temperatures $T_{e2}$ and $T_{i2}$, 
and magnetic field $B_{y02}$. 

We take 
the simulation domain in the $x$-$y$ plane 
and assume a perpendicular shock (i.e., $B_{x0}=0$). 
Since the ambient magnetic field is taken in the $y$ direction, 
free motion of particles along the ambient magnetic field 
is taken into account. 
As a motional electric field, a uniform external electric field 
$E_{z0} =-u_{x1}B_{y01} =-u_{x2}B_{y02}$ 
is applied in both upstream and downstream regions, 
so that both electrons and ions drift in the $x$ direction. 
At the left boundary of the simulation domain in the $x$ direction,
we inject plasmas with the same quantities 
as those in the upstream region, 
while plasmas with the same quantities as those 
in the downstream region are also injected from the right boundary 
in the $x$ direction.
We adopted absorbing boundaries 
to suppress non-physical reflection of electromagnetic waves at 
both ends of simulation domain in the $x$ direction \citep{Umeda_2001}, 
while the periodic boundaries are imposed 
in the $y$ direction.

The initial downstream quantities 
are given by solving 
the shock jump conditions for 
a magnetized two-fluid isotropic plasma
consisting of electrons and ions 
\citep[e.g.,][]{Hudson_1970}, 
\begin{eqnarray}
 \omega_{pe1}^2 u_{x1} = \omega_{pe2}^2 u_{x2}, \\
 \omega_{ce1} u_{x1} = \omega_{ce2} u_{x2}, \\
\omega_{pe1}^2 \left[ (1+\frac{m_i}{m_e}) u_{x1}^2 
+ (1+\frac{T_{i1}}{T_{e1}}) v_{te1}^2 \right]
+\frac{1}{2} \omega_{ce1}^2 c^2 
= \nonumber \\
 \omega_{pe2}^2 \left[ (1+\frac{m_i}{m_e}) u_{x2}^2 
+ (1+\frac{T_{i2}}{T_{e2}}) v_{te2}^2 \right]
+\frac{1}{2} \omega_{ce2}^2 c^2 , \\
\frac{1}{2} \omega_{pe1}^2 u_{x1}
\left[ (1+\frac{m_i}{m_e}) u_{x1}^2 +
 5(1+\frac{T_{i1}}{T_{e1}}) v_{te1}^2 \right] 
+ \omega_{ce1}^2 u_{x1} c^2 = \nonumber \\
\frac{1}{2} \omega_{pe2}^2 u_{x2} 
\left[ (1+\frac{m_i}{m_e}) u_{x2}^2 
+ 5(1+\frac{T_{i2}}{T_{e2}}) v_{te2}^2 \right] 
+ \omega_{ce1}^2 u_{x2} c^2,
\end{eqnarray}
where $T_s\equiv m_s v_{ts}^2$. 
In order to determine a unique initial downstream state, 
we need given upstream quantities 
$u_{x1}$, $\omega_{pe1}$, $\omega_{ce1}$, $v_{te1}$, and 
$T_{i1}/T_{e1}$ 
and an additional parameter. 
We assume a low-beta and weakly-magnetized plasma 
such that $\beta_{e1}=\beta_{i1}=0.125$ and 
$\omega_{ce1}/\omega_{pe1}=-0.1$ in the upstream region. 
We also use a reduced ion-to-electron mass ratio $m_i/m_e = 25$ 
for computational efficiency. 
The light speed $c/v_{te1}=40.0$ and 
the bulk flow velocity of the upstream plasma $u_{x1}/v_{te1}=4.0$ 
are also assumed.
Then, the Alfv\'{e}n Mach number is calculated as
$M_A = (u_{x1}/c)|\omega_{pe1}/\omega_{ce1}|\sqrt{m_i/m_e}=5.0$. 
The ion-to-electron temperature ratio 
in the upstream region is given as $T_{i1}/T_{e1}=1.0$. 
In this study, downstream ion-to-electron temperature ratio 
$T_{i2}/T_{e2} = 8.0$ is also assumed as 
another initial parameter 
to obtain the unique downstream quantities 
by solving the above four equations, 
$\omega_{pe2}/\omega_{pe1} = 1.8372$, 
$\omega_{ce2}/\omega_{pe1} = 0.3375$, 
$u_{x2}/v_{te1} = 1.1851$, 
and $v_{te2}/v_{te1} = 2.6393$.

In this study, we perform two runs with different 
sizes of the simulation domain. 
We use $N_x \times N_y = 2048 \times 1024$ 
cells for the upstream region and 
$N_x \times N_y = 2048 \times 1024$ 
cells for the downstream region, respectively, 
in Run A. 
The grid spacing and time step of the present simulation are 
$\Delta x/\lambda_{De1} = 1.0$ and $\omega_{pe1}\Delta t=0.0125$, 
respectively. 
Here $\lambda_{De1}$ is the electron Debye length upstream.
Thus the total size of the simulation domain is 
$10.24\lambda_i \times 5.12\lambda_i$ 
which is long enough to include the ion-scale rippled structure, 
where $\lambda_i = c/\omega_{pi1}(=200\lambda_{De1})$ is the 
ion inertial length. 
In Run B, we use $N_x \times N_y = 2048 \times 128$ 
cells for the upstream region and 
$N_x \times N_y = 2048 \times 128$ 
cells for the downstream region, respectively. 
Thus the the total size of the simulation domain is 
$10.24\lambda_i \times 0.64\lambda_i$, 
in which ion-scale processes along the ambient magnetic field 
is neglected. 
We used 16 pairs of electrons and ions per cell in the upstream region 
and 64 pairs of electrons and ions per cell in the downstream region, 
respectively, at the initial state.

\section{Result}

Figure 1 shows the transverse magnetic field $B_y$ 
as a function of position $x$ and time $t$ 
averaged over the $y$ direction for Run A. 
The position and time are renormalized by 
the ion inertial length $\lambda_i$ and 
the ion cyclotron period $1/\omega_{ci1}$, 
respectively. 
The magnitude is 
normalized by the initial upstream magnetic field $B_{y01}$. 
In the present shock-rest-frame model, 
a shock wave is excited 
by the relaxation of the two plasmas with different quantities. 
Figure 1 shows that the shock front 
appears and disappears at a timescale of the downstream 
ion gyro-period, which corresponds to the 
cyclic reformation of perpendicular shocks. 
Since the initial state is given by the shock jump conditions 
for a two-fluid plasma consisting of electrons and ions, 
the excited shock is ``almost'' at rest in the simulation domain. 

We have also performed a 1D simulation 
with the same parameter as the 2D simulations 
and confirmed that the period of cyclic reformation 
in all simulation runs (Runs A, B, and 1D) is the same. 
However, the peak amplitude of the overshooting magnetic field 
in Run A is about $6.0B_{y01}$, 
while this is about $4.5B_{y01}$ in Run B and in the 1D run. 
We have also confirmed that 
the cyclic reformation becomes less significant for $\omega_{ci1} t >7$ 
in Run A, 
while the cyclic reformation continues for a long time 
in Run B and in the 1D run.
The present result is in agreement with 
the recent 2D full-particle simulations of a low-Mach number shock, 
in which it has been demonstrated that the perpendicular shock 
evolves from the cyclic reformation phase 
to the ``nonlinear whistler'' phase 
in 2D simulations with the in-plane magnetic field 
\citep{Hellinger_2007,Lembege_2008}.

\begin{figure}
\includegraphics[width=0.5\textwidth]{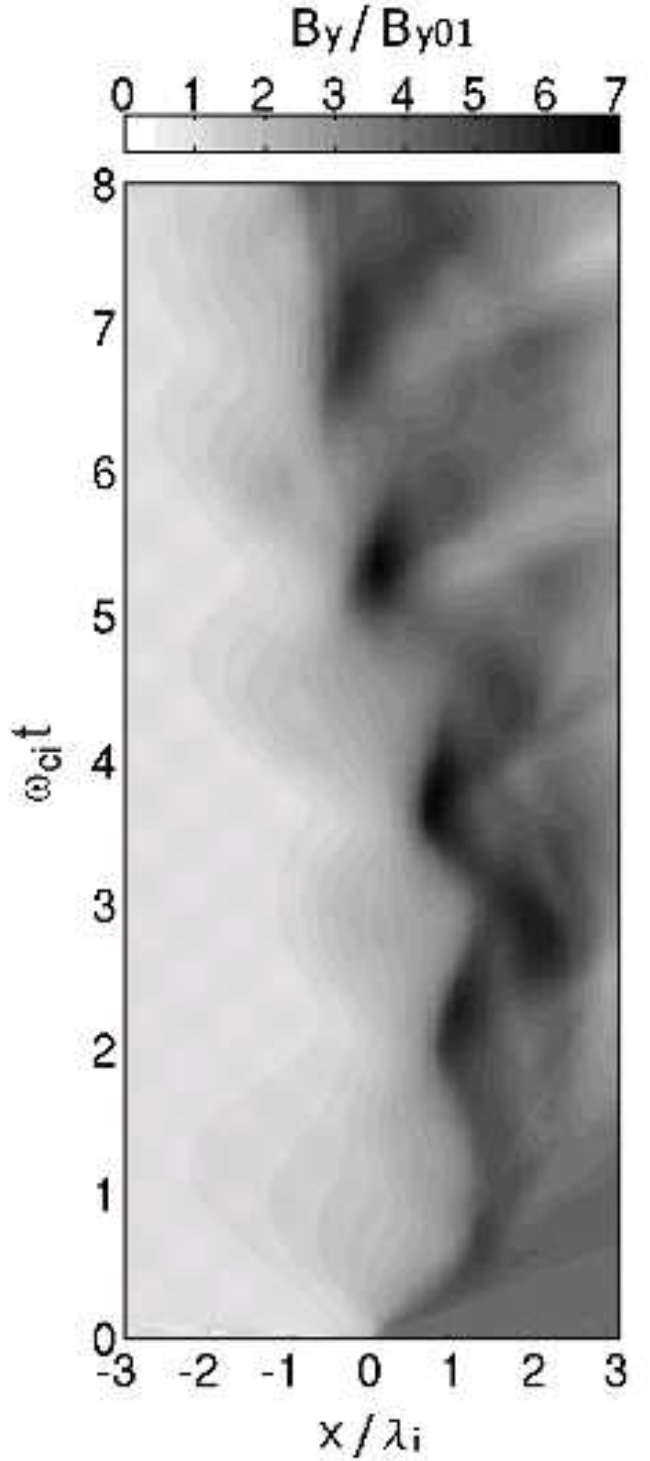}
\caption{
Transverse magnetic field $B_y$ 
as a function of position $x$ and time $t$ 
averaged over the $y$ direction for Run A. 
The position and time are normalized by 
$\lambda_i$ and $1/\omega_{ci1}$, 
respectively. 
The magnitude is 
normalized by the initial upstream magnetic field $B_{y01}$. 
}
\end{figure}


In the present study we focus on the 
particle acceleration in the cyclic reformation phase. 
Figure~2 shows the rippled structure 
of the perpendicular shock at $\omega_{ci1} t = 6.256$. 
The top panel shows a gray-scale map 
of the magnetic field magnitude $B_y$, 
and the bottom panel shows a gray-scale map 
of the electric field magnitude $E_x$ 
around the shock transition region. 
The magnitude of magnetic field is 
normalized by the initial upstream magnetic field $B_{y01}$, 
while the magnitude of electric field is
normalized by the motional electric field $E_{z0}$. 

We found a strong fluctuation in the magnetic field component 
$B_y$ at the shock surface. 
The amplitude of the fluctuation is estimated as 
$\sim 1.5B_{y01}$, which is larger than the magnitude of  
the upstream magnetic field. 
The wavelength of the fluctuation is several ion inertial length. 
This strong fluctuation has been analyzed by \citet{Lowe_2003} 
in terms of ripples at the shock overshoot. 

We computed a numerical frequency-wavenumber spectrum 
by taking Fourier transformation of the $B_z$ magnetic field component 
(not shown) and obtained a similar spectrum to 
the result by \citet{Lowe_2003}. 
The generation mechanism of the ripples is thought to be 
the ion perpendicular temperature anisotropy 
in the shock transition region, 
which drives Alfven ion cyclotron (L-mode) or mirror mode waves. 
Note that there is not any rippled structure in Run B 
because the size of the simulation domain 
in the magnetic field ($y$) direction 
is shorter than the ion inertial length. 
The overshoot magnetic field $B_y$ is uniform along the magnetic field 
in Run B 
and its magnitude is about $4.5B_{y01}$, which is equal to 
the average overshoot magnetic field in Run A.

Around the shock front ($x/\lambda_i = -0.5 \sim 0$), 
we also found a strong negative electric field 
with a magnitude of $\sim -4.0E_{z0}$
(see the bottom panel of Figure 2). 
This strong negative electric field at the shock front 
reflects ions upstream. 
Such a negative electric field is also found in Run B, 
but its magnitude is much smaller, about $E_x/E_{z0} \sim -2.0$. 
There is a good correlation between the structure of 
$B_y$ and $E_x$ at the shock front as seen in Figure 2, 
implying that the strong negative electric field is 
associated with the rippled structure.

\begin{figure}
\includegraphics[width=0.5\textwidth]{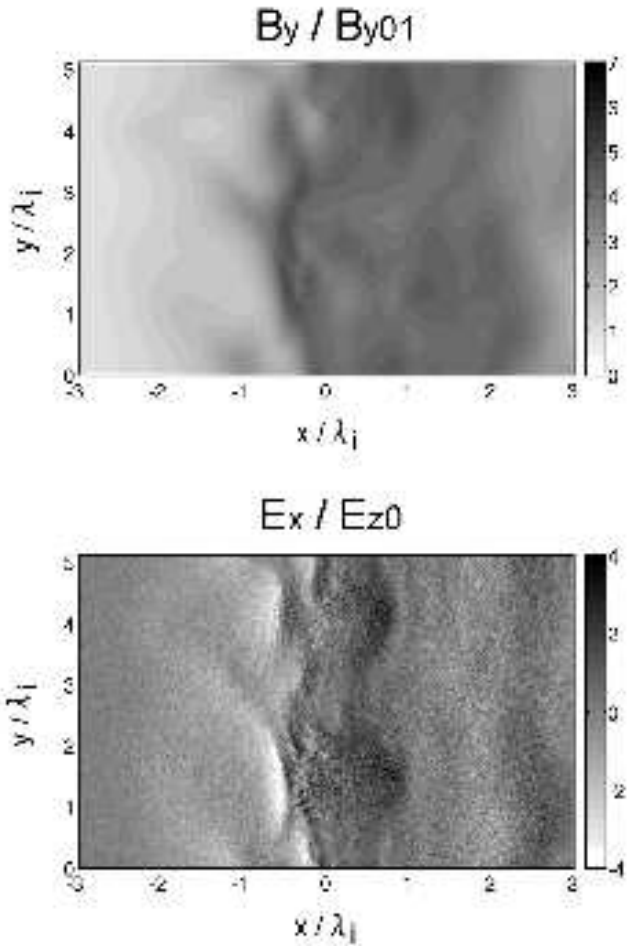}
\caption{
The rippled structure around the transition region 
of the perpendicular shock 
at $\omega_{ci1}t = 6.256$ for Run A.  
The magnetic field magnitude $B_y$ (top) 
and the electric field magnitude $E_x$ (bottom). 
The magnitude of magnetic field and the electric field is 
normalized by the initial upstream magnetic field $B_{y01}$, 
and the motional electric field $E_{z0}$, respectively. 
\label{fig2}
}
\end{figure}


\begin{figure*}
\includegraphics[width=1.0\textwidth]{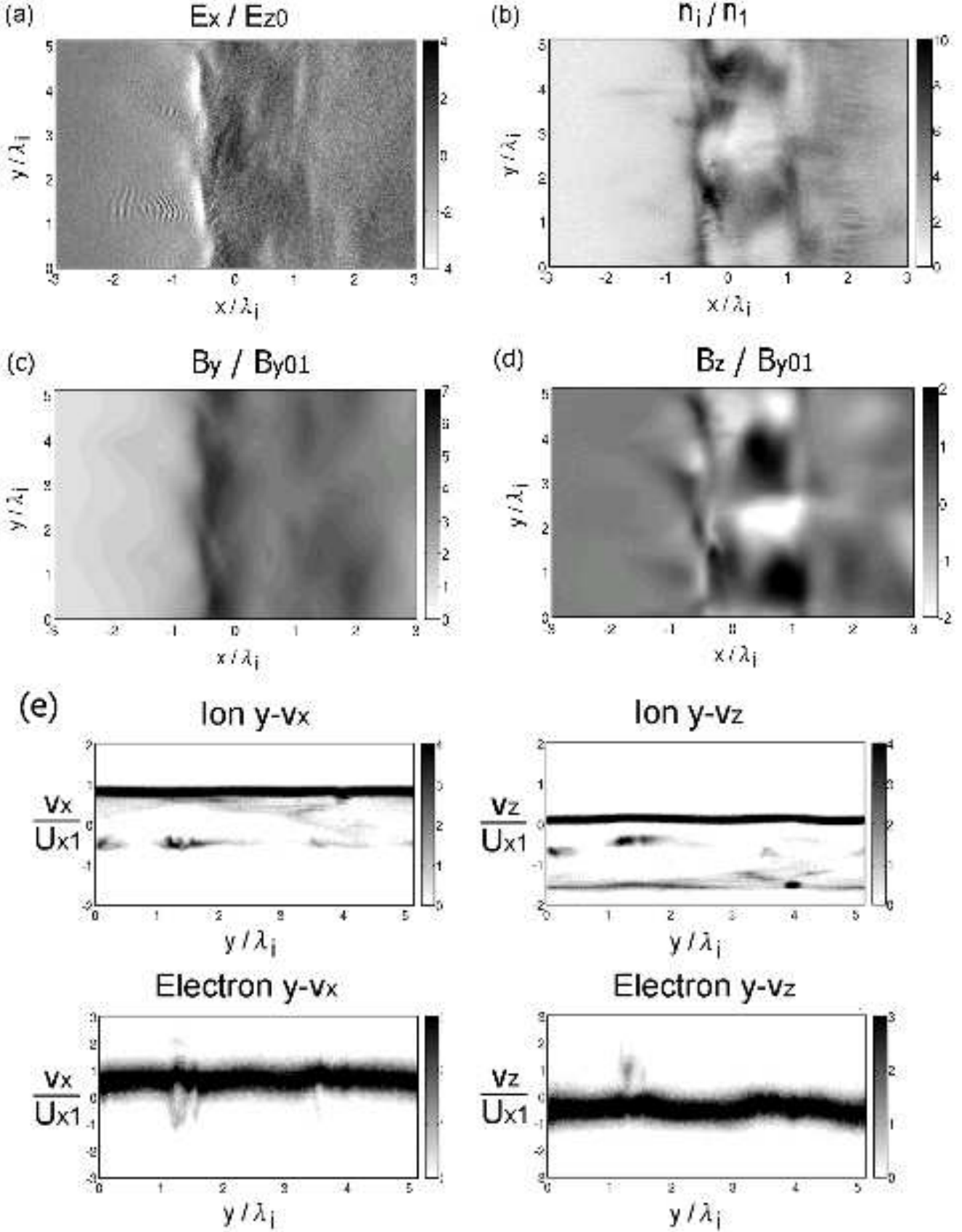}
\caption{
(a) The electric field magnitude $E_x$ 
at $\omega_{ci1}t = 6.575$ for Run A.  
(b) The corresponding ion density $n_i$. 
(c) The corresponding magnetic component $B_y$. 
(d) The corresponding magnetic component $B_z$. 
(e) The corresponding reduced phase-space distribution functions in 
$y-v_x$ and $y-v_z$, 
at $x/\lambda_i = -1.0$. 
The density is normalized by the upstream density $n_1$. 
The velocity is normalized by the upstream bulk velocity $u_{x1}$. 
}
\end{figure*}

The effect of the acceleration of reflected ions 
is analyzed in Figure 3. 
The electric field magnitude $E_x$ 
at $\omega_{ci1}t = 6.575$ for Run A is shown in Figure 3a. 
At $(x/\lambda_i,y/\lambda_i) \sim$ 
(-1.0,1.5) and (-1.0,3.5), 
we found wave structures with a short wavelength, 
which corresponds to electrostatic waves 
excited by the current-driven instability due to 
reflected ions. 
An interesting result here is that 
the electrostatic waves are not excited uniformly 
in the $y$ direction 
but are excited in localized regions. 
It is suggested that 
the localized excitation of the electrostatic waves 
is due to the strong negative shock-normal electric field 
at the shock front 
that accelerates reflected ions upstream. 

The strong negative shock-normal electric field 
is associated with the rippled strucuture. 
Figures 3c and 3d show the mangetic field components 
$B_y$ and $B_z$ at $\omega_{ci1}t = 6.575$ for Run A. 
We have analyzed the Hall electric field 
$(\vec{J} \times \vec{B})/en$ 
and found that the term of $J_z B_y/en$ is dominant. 
In other words, the strong negative shock-normal electric field 
is due to 
the magnetic pressure gradient force of the $B_y$ componet 
around $x/\lambda_i \sim -0.5$ arising from the rippled structure.  
Here, the excitation of electrostatic waves is discussed 
in terms of the particle distributions 
shown in Figures 3b and 3e. 

Figure 3b shows the ion density $n_i$ 
normalized by the upstream density $n_1$. 
Around $y/\lambda_i \sim$ 0.2, 2.0, and 3.2, 
we found the enhancement of the ion density that 
exceeds $15n_1$. 
In these regions, the magnitude of the magnetic field 
becomes lower (see $B_y$ in Figure 3c)
to keep the pressure balance.

The left panels of Figure 3e show $y-v_x$ phase-space plots of 
ions and electrons. 
Around $y/\lambda_i \sim$ 0, 1.5, and 3.5, 
we found ion components with negative velocity, 
which correspond to reflected ions. 
The electrostatic waves around 
$y/\lambda_i \sim$ 1.5 and 3.5 
are excited by the localized reflected ions. 
The patchy reflected ion beams are formed by 
the localized negative shock-normal electric field 
around $x/\lambda_i = -0.5$.

Around $y/\lambda_i \sim$ 0, 
despite a strong shock-normal electric field,
electrostatic waves are not strongly excited, either. 
This is because of small number of ions. 
We need a strong ion current in the $-x$ direction 
to enhance the current-driven instability. 
That is, the velocity of reflected ions should be high 
and/or the number of reflected ions should be high 
to excite the patchy electrostatic waves.


In the $y-v_z$ phase-space plots in the right panels, 
nonthermal components of electrons 
are found at $y/\lambda_i \sim$ 1.5, 
where electrostatic waves are strongly enhanced. 
The existence of nonthermal components in $v_z$ 
is the evidence of the electron surfing acceleration 
at low-Mach-number perpendicular shock with $M_A=5$. 
The electron surfing acceleration 
takes place in a localized region 
where a negative shock-normal current due to 
reflected ions is strongly enhanced.


\begin{figure}[t]
\includegraphics[width=0.5\textwidth]{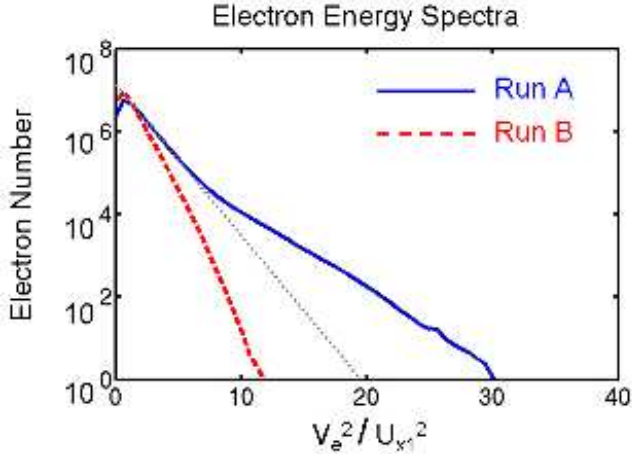}
\caption{
Energy distribution functions of electrons 
in the downstream region. 
The solid line shows the result of Run A 
and the dashed line shows the result of Run B. 
The energy is normalized by the upstream bulk energy of electrons. 
The dotted line indicates a Maxwellian distribution 
for the downstream of Run A. 
}
\end{figure}

Figure 4 shows energy distribution functions of electrons 
in the downstream region 
for Runs A and B. 
The solid line shows the result of Run A, 
and the dashed line shows the result of Run B. 
Firstly, we found nonthermal electrons in Run A 
while there is no nonthermal electrons in Run B. 
Secondly, the downstream temperature in Runs A and B 
is very different, indicating 
the existence of additional electron energization mechanism 
in Run A, i.e., rippling of the perpendicular shock 
discussed by \citet{Burgess_2006b}. 
We have also performed a 1D simulation run and 
confirmed that the electron energy spectrum is 
essentially the same as that in Run B. 

The electron energy distribution for Run B 
(dashed line) corresponds to 
the Maxwellian distribution function with 
thermal velocity of $v_{te}\sim 2.3$, 
which can be explained by 
the plasma heating by magnetic compression. 
On the other hand, 
the thermal component of electron energy distribution for Run A 
(dotted line) corresponds to
the Maxwellian distribution function with 
thermal velocity of $v_{te}\sim 3.1$. 
In addition, the maximum energy of electrons reaches $\sim 30u_{x1}^2$ 
in Run A. 

The rippled structure can further energize 
electrons as discussed by \citet{Burgess_2006b}, 
which can explain the Maxwellian distribution for Run A. 
In the shock foot region ($x/\lambda_i = -1\sim2$), 
the magnitude of the magnetic field is $B_y \sim 1.5B_{y01}$. 
The average amplitude of electrostatic waves excited by the 
current-driven instability is $E_x \sim 3E_{z0}$. 
Thus the maximum energy of non-thermal electrons 
becomes about $4u_{x1}^2 \sim <v_x^2+v_z^2> = |E_x/B_y|^2$ 
via the electron surfing acceleration 
as indicated from the $y-v_x$ and $y-v_z$ phase-space plots 
shown in Figure 3e. 
Thus there is three effects for energization of non-thermal electrons: 
The first is the electron surfing acceleration; 
The second is the magnetic compression at the shock overshoot 
\citep[e.g.,][]{Shimada_2000,Schmitz_2002a,Umeda_2006}. 
The third is the scattering by ion-scale rippled structure
\citep{Burgess_2006b}. 
To reach the maximum energy of $\sim 30u_{x1}^2$ 
all the three effects are needed.

Both in Run B and in the 1D run, 
the current-driven instability 
due to reflected ions excites electrostatic waves
with amplitude of $\sim 2.0 E_{z0}$. 
However, this amplitude is not enough 
for the electron surfing acceleration. 
By contrast, 
the amplitude of the locally-excited electrostatic wave 
exceeds $\sim 4.0 E_{z0}$ in Run A as shown in Figure 3a, 
and energetic electrons are generated 
by the electron surfing acceleration. 
This result suggests that microscopic 
current-driven instability enhanced by 
the ion-scale rippled structure 
of the low-Mach-number perpendicular shock 
generates nonthermal electrons, 
as observed in moderate-Mach-number perpendicular shocks 
of $M_A = 10-20$ 
\citep[e.g.][]{Shimada_2000,Hoshino_2002,
Schmitz_2002a,Schmitz_2002b,Amano_2007}.

\section{Summary and Discussion}

We have studied electron acceleration at a low-Mach-number 
perpendicular collisionless shock by performing 
two-dimensional full particle simulations. 
In order to take into account the effect of 
the rippling of a perpendicular shock 
\citep{Lowe_2003,Burgess_2006a,Burgess_2006b}, 
the simulation domain is taken to be larger than the 
ion inertial scale by using the shock-rest-frame model
\citep{Umeda_2008}. 
In the previous works, 
the electron acceleration by electron-scale microscopic instabilities 
at low-Mach-number perpendicular shocks has not focused on 
because the electron surfing acceleration by electrostatic waves 
\citep[e.g.,][]{Shimada_2000,Hoshino_2002}, 
which is one of the possible injection mechanisms for electrons, 
is thought to be effective only in high-Mach-number perpendicular shocks. 
The present result suggests, by contrast, that 
the electron surfing acceleration is also a common feature 
at a perpendicular shock with $M_A = 5$. 

The mechanism for generation of high-energy electrons 
is quite simple, and is summarized as follows:  
The perpendicular shock forms the rippled structures  
by ion temperature anisotropy; 
The rippled structures excite a 
strong electric field component in the shock-normal direction; 
Reflected ions in the shock transition region is 
strongly accelerated upstream by the electric field component 
of rippled structures; 
The strongly-accelerated reflected ions 
give a high growth rate of a current-driven instability 
to large-amplitude electrostatic waves 
in a localized region; 
Energetic electrons are generated by 
the electrostatic waves via surfing acceleration. 
Finally, high-energy non-thermal electrons are 
generated by both magnetic compression 
and scattering by ion-scale rippled structure 
in the shock transition region. 

It has been demonstrated that multi-dimensionality sometimes weakens 
the electron surfing acceleration 
in the 2D simulations of 
uniform plasma models \citep[e.g.,][]{Dieckmann_2006,Ohira_2007}
and self-consistent shock models 
\citep[e.g.,][]{Dieckmann_2008,Umeda_2008}. 
Very recently, however, \citet{Amano_2008} 
have found the electron surfing acceleration
in 2D simulation of a perpendicular shock 
with out-of-plane magnetic field. 
In their 2D simulation, the simulation domain is taken to be 
much larger than the ion inertial length along the 
shock surface, and kinetic effects of ions are fully included. 
The present result is another example showing the 
effect of ion kinetics to the electron surfing acceleration. 
It has been confirmed that 
the cross-scale coupling between 
an ion-scale mesoscopic instability 
and an electron-scale microscopic instability 
is important. 
Hence, a large-scale full particle simulation would be 
essential for studies of 
the electron acceleration at collisionless shocks.

Finally let us discuss on the reduced mass ratio. 
In the present simulation with $M_A=5$, $\omega_{pe1}/\omega_{ce1}=10$, 
and $\beta = 0.125$, the Buneman-type mode becomes unstable 
at $\omega_{UHR}\simeq \omega_{pe1}$ when we use a reduced mass ratio 
of $m_i/m_e=25$. 
When we use the real mass ratio of $m_i/m_e=1836$, 
the thermal velocity of upstream electrons becomes 
about 8.6 times as large as the case of $m_i/m_e=25$, 
the self-reformation process is suppressed 
and the Buneman-type mode is also stabilized \citep{Scholer_2004}. 
To make the Buneman-type mode unstable 
the electron thermal velocity must be smaller than the 
relative velocity between incoming electrons and reflected ions. 
Thus the present simulation result can be applied 
to a much lower beta of $\beta < 0.03$.

\acknowledgments

The authors are grateful to Y.~Ohira and Y.~Fujita
for helpful comments. 
The computer simulations were carried out 
on Fujitsu HPC2500 at ITC, Nagoya University, and 
NEC SX-7 at YITP, Kyoto University, 
as a collaborative computational research project at 
STEL, Nagoya University, and 
YITP, Kyoto University. 
This work was supported by Fujii research grant 
from Hiroshima University, Grant-in-aid 18740153, 19047004 (R. Y.), 
and in part 17GS0208 (T. U.) from MEXT of Japan.


\begin{thebibliography}{}

\bibitem[Amano \& Hoshino(2007)]{Amano_2007}
Amano, T., \& Hoshino, M. 2007, 
\apj, 661, 190

\bibitem[Amano \& Hoshino(2008)]{Amano_2008}
Amano, T., \& Hoshino, M. 2008, 
\apj, in press

\bibitem[Bamba et al.(2003)]{bamba2003}
Bamba, A., Yamazaki, R., Ueno, M., and Koyama, K. 2003, \apj, 589, 827

\bibitem[Bamba et al.(2005)]{bamba2005}
Bamba, A., et al. 2005, \apj, 621, 793

\bibitem[Blandford \& Eichler(1987)]{blandford1987}
Blandford, R. D. \& Eichler, D. 1987, Phys. Rep. 154, 1

\bibitem[Brogan et al.(2006)]{brogan2006}
Brogan, C. L. et al. 2006, ApJ, 639, L25

\bibitem[Burgess(2006a)]{Burgess_2006a}
Burgess, D. 2006a, 
\jgr, 111, A10210 

\bibitem[Burgess(2006b)]{Burgess_2006b}
Burgess, D. 2006b, \apj, 653, 316

\bibitem[Bykov et al.(2000)]{bykov2000}
Bykov, A. M. et al. 2000, ApJ, 538, 203

\bibitem[Claussen et al.(1997)]{claussen1997}
Claussen, M. J. et al. 1997, ApJ, 489, 143

\bibitem[Chevalier(1999)]{chevalier1999}
Chevalier, R. A. 1999, ApJ, 511, 798

\bibitem[Dieckmann \& Shukla(2006)]{Dieckmann_2006}
Dieckmann, M., \& Shukla, P. K. 2006, 
Plasma Phys. Controll Fusion, 48, 1515. 

\bibitem[Dieckmann et al.(2008)]{Dieckmann_2008}
Dieckmann, M. E., Meli, A., Shukla, P. K., 
Drury, L. O. C. \& Mastichiadis, A. 2008, 
Plasma Phys. Controll Fusion, 50, 065020. 

\bibitem[Drury(1983)]{drury1983}
Drury, L. O'C. 1983, Rep. Prog. Phys. 46, 973

\bibitem[Fujita et al.(2007)]{fujita2007}
Fujita, Y. et al. 2007, ApJ, 663, L61

\bibitem[Hellinger et al.(2007)]{Hellinger_2007}
Hellinger, P., Travnicek, P., Lembege, B., \& Savoini P. 2007, 
\grl, 34, L14109 


\bibitem[Hoshino \& Shimada(2002)]{Hoshino_2002}
Hoshino, M. \& Shimada, N. 2002, 
\apj, 572, 880

\bibitem[Hudson(1970)]{Hudson_1970}
Hudson, P. D. 1970, 
\planss, 18, 1611

\bibitem[Katsouleas \& Dawson(1983)]{Katsouleas_1983}
Katsouleas, T., \& Dawson, J. M. 1983, 
Phys. Rev. Lett., 51, 392

\bibitem[Koyama et al.(1995)]{koyama1995}
Koyama, K. et al. 1995, Nature, 378, 255

\bibitem[Lee et al.(2004)]{Lee_2004}
Lee, R. E., Chapman, S. C., \& Dendy, R. O. 2004, 
\apj, 604, 187

\bibitem[Lembege et al.(2008)]{Lembege_2008}
Lembege, B., Savoini, B., Hellinger, P., \& Travnicek, P. M. 2008, 
\jgr, in press

\bibitem[Leroy et al.(1981)]{Leroy_1981}
Leroy, M. M., Goodrich, C. C., Winske, D., 
Wu, C. S. \& Papadopoulos, K. 1981, 
\grl, 8, 1269

\bibitem[Leroy et al.(1982)]{Leroy_1982}
Leroy, M. M., Winske, D., Goodrich, C. C., 
Wu, C. S. \& Papadopoulos, K. 1982, 
\jgr, 87, 5081

\bibitem[Lowe \& Burgess(2003)]{Lowe_2003}
Lowe, R. E., \& Burgess, D. 2003,
Ann. Geophys. 21, 671

\bibitem[Markevitch et al.(2002)]{markevitch2002}
Markevitch, M. et al. 2002, ApJ, 567, L27

\bibitem[McClements et al.(2001)]{McClements_2001}
McClements, K. G., Dieckmann, M. E., Ynnerman, A., 
Chapman, S. C., \& Dendy, R. O. 2001, 
\prl, 87, 255002


\bibitem[Ohira \& Takahara(2007)]{Ohira_2007}
Ohira, Y., \& Takahara, F. 2007, 
\apjl, 661, L171

\bibitem[Oka et al.(2006)]{oka2006}
Oka, M. et al. 2006, \grl, 33, L24104

\bibitem[Schmitz et al.(2002a)]{Schmitz_2002a}
Schmitz, H., Chapman, S. C., \& Dendy, R. O. 2002a,
\apj, 570, 637

\bibitem[Schmitz et al.(2002b)]{Schmitz_2002b}
Schmitz, H., Chapman, S. C., \& Dendy, R. O. 2002b,
\apj, 579, 327

\bibitem[Scholer \& Matsukiyo(2004)]{Scholer_2004}
Scholer, M., \& Matsukiyo, S. 2004, 
Ann. Geophys. 22, 2345

\bibitem[Shimada \& Hoshino(2000)]{Shimada_2000}
Shimada N., \& Hoshino, M. 2000, 
\apjl, 543, L67

\bibitem[Umeda(2004)]{Umeda_PhD}
Umeda, T. 2004, 
Ph.D. Thesis, Kyoto University

\bibitem[Umeda et al.(2001)]{Umeda_2001}
Umeda, T., Omura, Y., \& Matsumoto, H. 2001, 
Comput. Phys. Commun., 137, 286

\bibitem[Umeda et al.(2003)]{Umeda_2003}
Umeda, T., Omura, Y., Tominaga, T., \& Matsumoto, H. 2003, 
Comput. Phys. Commun., 156, 73

\bibitem[Umeda \& Yamazaki(2006)]{Umeda_2006}
Umeda, T., \& Yamazaki, R. 2006, 
Earth Planets Space, 58, e41
(arXiv:physics/0607220)

\bibitem[Umeda et al.(2008)]{Umeda_2008}
Umeda, T., Yamao, M., \& Yamazaki, R. 2008, 
\apjl, 681, L85

\bibitem[Winske \& Quest(1988)]{Winske_1988}
Winske, D., \& Quest, K. B. 1988, 
\jgr, 93, 9681

\bibitem[Yamazaki et al.(2006)]{yamazaki2006}
Yamazaki, R. et al. 2006, MNRAS, 371, 1975

\end{thebibliography}
\end{document}